\documentclass[11pt, letterpaper]{article}
\usepackage{tcolorbox}     
\usepackage[utf8]{inputenc}
\usepackage[T1]{fontenc}
\usepackage{amsmath, amssymb}
\usepackage[margin=1in]{geometry}
\usepackage[colorlinks=true, urlcolor=blue, linkcolor=blue, citecolor=blue]{hyperref}
\usepackage{graphicx} 
\usepackage{subcaption} 
\usepackage{lipsum} 
\usepackage{tikz}
\usepackage{booktabs}
\usepackage{tcolorbox}
\usepackage[explicit]{titlesec} 
\usepackage{tabularx}
\usepackage{wrapfig}

\usepackage[left,modulo]{lineno}

\newcommand{\vignettetitle}[1]{\par\textbf{#1}\par\nobreak}

\newcommand{\footremember}[2]{%
    \footnote{#2}
    \newcounter{#1}
    \setcounter{#1}{\value{footnote}}%
}

\usepackage{epigraph}

\usetikzlibrary{
    shapes.geometric, 
    arrows.meta,      
    positioning,      
    calc,              
    intersections,     
    fit              
}

\usepackage[backend=bibtex,style=numeric-comp,sorting=none]{biblatex}
\title{\textbf{The Variance Paradox: How AI Reduces Diversity but Increases Novelty}}
\bigskip
\author{Bijean Ghafouri\footremember{alley}{bghafour@usc.edu} \\
University of Southern California\\
}

\begin{document}
\date{}
\maketitle

\begin{abstract}
The diversity of human expression is the raw material of discovery. Generative artificial intelligence threatens this resource even as it promises to accelerate innovation, a paradox now visible across science, culture, and professional work. We propose a framework to explain this tension. AI systems compress informational variance through statistical optimization, and users amplify this effect through epistemic deference. We call this process the AI Prism. Yet this same compression can enable novelty. Standardized forms travel across domain boundaries, lowering translation costs and creating opportunities for recombination that we term the Paradoxical Bridge. The interaction produces a U-shaped temporal dynamic, an initial decline in diversity followed by recombinant innovation, but only when humans actively curate rather than passively defer. The framework generates testable predictions about when compression constrains versus amplifies creativity. As AI becomes infrastructure for knowledge work, managing this dynamic is essential. Without intervention, the conditions for recovery may not arrive.
\end{abstract}

\newpage


\section*{Main}
The accumulation and recombination of knowledge have long driven human progress \cite{Campbell1960,arrow1962economic, popper2014conjectures, nelson1985evolutionary, simon1987steam, basalla1988evolution, Romer1990, david1990dynamo, Weitzman1998, dennett1996darwin, arthur2009nature, page2010diversity, mokyr2011gifts}. Yet the same technologies designed to accelerate the creation and spread of ideas can also narrow the diversity that fuels them. This tension—between informational convergence and creative exploration—has accompanied every major transformation in how humans produce, communicate and preserve knowledge. The rise of generative artificial intelligence amplifies this dynamic at unprecedented scale \cite{brinkmann2023machine}. 

Scholars across the behavioural and social sciences have examined the sources of novelty that drive cultural and intellectual evolution. Anthropologists trace this capacity to cumulative culture \cite{Boyd1985, Henrich2016}, historians to informational infrastructures \cite{Eisenstein1979, Grafton2021}, economists to knowledge as the engine of growth \cite{Romer1990, Weitzman1998}, psychologists to creativity as a universal resource \cite{Simonton1999, amabile2020creativity}, and sociologists to the opportunities created by social structures \cite{simmel2010conflict, merton1968social, burt2004structural}. These traditions converge on a common view: novelty arises from processes of variation, transmission, and recombination. Each new information technology has altered these processes, reshaping the conditions under which innovation emerges \cite{Innis1950, Eisenstein1979, Castells1996, Benkler2006, Muthukrishna2016}.

Generative AI continues this historical lineage while introducing a distinctive mechanism, informational compression, that transforms how ideas vary, spread, and combine. Large language models are designed to optimize for statistical averages, reproducing what is probable rather than what is distinctive. Compared to human expression, which tends to generate long-tailed distributions of ideas, language, and styles, these systems produce outputs that cluster more tightly around the mean. The result is a measurable convergence of linguistic, aesthetic, and conceptual forms across academia \cite{Anderson2024, Prakash2025, Leppanen2025, sourati2025shrinking}, professional work \cite{Noy2023, OECD2025, DataInnovation2024, Song2024}, education \cite{Doshi2024, Padmakumar2023}, and creative production \cite{Boden2004, Zhou2024, Vaithilingam2022, sarkar2023exploring, Homogenizing2025, Bender2021}. This raises a central question: if innovation depends on variation, what happens when the tools of discovery are engineered to reduce it? We propose a framework for studying how AI-driven compression can both constrain and enable creativity through new forms of recombination, and outline a research agenda for investigating these dynamics.

However, the story is not one of simple convergence. A growing body of research shows that AI can also serve as a catalyst for novelty, particularly when humans act as curators or when algorithms link ideas across domains \cite{gao2023quantifying, LeeChung2024, Shiiku2025, Krakowski2025, Johansen2025, zhou2025expands}. These seemingly contradictory effects—compression and creativity—present a central paradox. The same intelligent systems that narrow variation can also expand the space of possible ideas. This paradox is best understood through recombination, the basic mechanism of innovation by which new forms emerge from rearranging existing elements \cite{Schumpeter1942, Campbell1960, Boden2004, Simonton1999, Weitzman1998, Arthur2009, Padgett2012, Muthukrishna2016}. Our question is not whether AI suppresses or enhances creativity, but when and how compression becomes a precondition for recombination.

Historical periods of informational narrowing have often created the conditions for later bursts of novelty. The typographical fixity of the printing press established a shared foundation for the Scientific Revolution \cite{Eisenstein1979, Shapin1996, Grafton2021}, and the methodological constraints of early science fostered the disciplined accumulation of new insights \cite{Kuhn1962, Latour1987}. Standardized general-purpose technologies, from interchangeable parts to the internet, have similarly led to large-scale recombination in economics and communication \cite{Bresnahan1995, Helpman1998, mokyr2011gifts}. The question now is whether generative AI, with its unprecedented scale and speed, will follow a similar evolutionary path, transforming temporary homogenization into global innovation.

To address this question, we propose a framework that links these opposing forces within a single explanatory structure. We describe how informational compression, which we call the AI Prism, interacts with creative recombination, modeled as a Paradoxical Bridge. Together, they explain when AI-driven homogenization becomes a dead end for creativity, and when it becomes a resource for cross-domain discovery. In this view, simplification is not always the enemy of progress: under certain conditions, local uniformity can be a necessary precondition for global novelty.

\vspace{0.5cm}
\begin{tcolorbox}[title=\normalsize Summary of our Contributions, colback=white,colframe=black]
\begin{itemize}\itemsep0.25em
\item \textbf{Unified Mechanism:} We link two literatures usually treated apart—variance compression (homogenization) and cross-domain connection (novelty)—into a single \textit{Prism–Bridge} framework.
\item \textbf{Human Substrate:} We introduce the concept of \textit{AI-derivative epistemology}, describing a mode of knowing in which users treat an AI’s output as self-sufficient evidence rather than a starting point for inquiry.
\item \textbf{Testable Prediction:} We propose a U-shaped temporal dynamic, arguing that an initial decline in variance is a precondition for a subsequent rise in recombinant novelty. Our framework specifies the conditions and turning points that govern this process.
\item \textbf{Governance Levers:} We outline how policy, design, and education can maintain equilibrium between uniformity and renewal, reframing debates about AI safety and creativity as problems of sustaining informational diversity.
\end{itemize}
\end{tcolorbox}
\vspace{1cm}

\section*{Homogenizing AI Prism}
Generative AI has a paradoxical architecture. Models are trained on immense heterogeneity drawn from the digital landscape, yet their objective is to compress this diversity into statistically probable outputs that fit prior patterns\footnote{To avoid terminological ambiguity, we treat informational diversity as a parent construct with three measurable dimensions. These are statistical variance, information-theoretic entropy, and semantic diversity, which capture distinct forms of dispersion like statistical spread and unpredictability. Following prior work in cultural evolution and information theory \cite{lin2002divergence, jost2006entropy, Muthukrishna2016}, we use variance and entropy as operational proxies for informational diversity, while recognizing they are not formally equivalent. Our framework thus defines variance compression as the statistical reduction of informational diversity, while homogenization is its sociocultural manifestation.}. The result is not an expansion of informational variance, but a systematic reduction of it \cite{Wu2024, Qiu2025a, Leppanen2025}. Normatively, this is a worrying outcome. The diversity of perspectives and ideas is the raw material of adaptive intelligence \cite{Henrich2004, Hong2004, Muthukrishna2016}, so as variance decreases, a system's capacity for resilience, novelty, and discovery also declines.

To conceptualize this process, we propose the metaphor of the \textit{AI Prism}. As a physical prism refracts light into a narrow band of colors, the AI Prism refracts human informational diversity into a reduced set of outputs. This convergence is a predictable consequence of two interacting forces: the technical optimization of large language models, and human epistemic deference, our tendency to accept fluent and effortless answers \cite{Messeri2024, Bender2021}.

\subsection*{Architecture of convergence}
The first half of the \textit{Prism} dynamic is the statistical architecture of generative models, a structural property that systematically reduces informational variance. This process begins with the training objective. Language models are optimized for next-token prediction, which assigns a higher probability to common sequences while treating rare patterns as noise \cite{brown2020language, radford2019language}. This mirrors concerns about ‘stochastic parroting’ where models reproduce central tendencies and marginalize outliers \cite{Bender2021, Suhara2020, Angelidis2021, Bakker2022, Small2023}. Variance compression is then amplified during alignment. Although Reinforcement Learning from Human Feedback (RLHF) makes models safer, it also creates a regression-to-the-mean effect by rewarding centrality in rater preferences, thus narrowing stylistic diversity \cite{casper2023open, Xie2025, Wu2024}. At the decoding layer, conservative sampling strategies, such as low-temperature settings, further reduce variation compared to human-authored text \cite{sourati2025shrinking}.

These internal mechanics are amplified by system-level broadcast dynamics, driven by institutional incentives that favor a few dominant foundation models \cite{Bommasani2022}. This centralization reinforces convergence, particularly through translation asymmetries that diffuse standardized idioms globally \cite{Nakadai2023, Moser2023}. Ultimately, these combined technical forces do not just produce content, but produce it in a specific way. The outputs consistently presented to users are fluent, statistically probable, and seemingly authoritative.

\begin{tcolorbox}[floatplacement=!ht, float,
    title=\textbf{Box 1 | Examples of the Prism–Bridge Framework},
    colback=white,
    colframe=black,
    fonttitle=\bfseries,
    sharp corners,
]
These vignettes illustrate how AI-driven homogenization (the Prism) can enable novel recombination (the Bridge) in a two-stage process across different domains.

\vspace{1em}
\hrule 
\vspace{1.5em}

\begin{tabularx}{\linewidth}{@{} X @{\hspace{3em}} X @{}}

\vignettetitle{Screenwriting}
AI story generators trained on successful screenplays learn the conventions of three-act structure and character archetypes. Consequently, AI-assisted scripts often exhibit predictable plot points, creating a uniform "house style" for mainstream film. This makes stories easier to pitch and package for a global market, yet it also allows for radical recombination. A writer can take the structure from a classic Western, populate it with characters from a science fiction epic, and use AI to blend the dialogue, creating a coherent hybrid genre. The standardized structure becomes the canvas for novelty.
&
\vignettetitle{Architectural Design}
AI design tools produce initial blueprints that converge on established styles. A prompt for a "modern sustainable home," for example, reliably generates a sleek, glass-and-wood structure. This standardization simplifies client presentations, but also creates a global "kit of parts." An architect in Seoul can now integrate a novel support system from a design first conceived in Berlin, because both share the same underlying stylistic language. The homogenization of form enables the recombination of function.
\\

\addlinespace[3em] 

\vignettetitle{Drug Discovery}
AI platforms for drug discovery analyze immense datasets to predict promising new compounds. Their models, optimized to identify candidates resembling successful drugs, lead research pipelines to converge on known biological pathways. This accelerates early-stage research, but also creates a standardized language for molecular properties. A team working on Alzheimer's might notice an AI-generated compound for their disease shares a functional subgroup with an AI-generated cancer therapeutic. This shared language allows them to recombine the two, targeting a newly discovered link between the fields.
&
\vignettetitle{Finding Information}
AI-driven search engines increasingly offer a single, synthesized summary instead of a list of links. Optimized for clarity, these summaries converge on a standardized explanation for complex topics like "inflation" or "photosynthesis." This narrows the immediate diversity of viewpoints, but also provides a shared factual baseline for millions. A student can then take this standardized definition and easily connect it to another, spotting a real-world link they might have missed in a sea of jargon. The homogenization of answers enables the recombination of knowledge.
\\

\end{tabularx}
\end{tcolorbox}

\subsection*{AI-derivative epistemology}

The psychological foundations of the AI Prism lie in a shift in how individuals seek and validate knowledge. Generative systems deliver convergent outputs that align with statistical centrality, and users increasingly treat these outputs as adequate substitutes for inquiry \cite{Bender2021, Wu2024}. We describe this shift as AI-derivative epistemology: the systematic delegation of knowledge formation to artificial systems and the acceptance of their outputs with limited critical revision. The change represents not only a new cognitive shortcut but also a reorganization of epistemic authority.

The mechanism begins with the effort–trust tradeoff. Human cognition is bounded, and individuals adopt strategies that minimize cognitive expenditure \cite{Hardwig1985, Kitcher1990}. Historically, this produced the division of cognitive labor in scientific and institutional settings \cite{Hutchins1995}. Generative AI extends that principle to everyday reasoning by providing immediate and fluent answers that reduce mental effort \cite{castro2023human, Liao2022}. The efficiency gained from this exchange increases trust in the system’s reliability and discourages secondary verification.

Fluency bias strengthens this dynamic. Large language models produce text that is coherent and polished but not necessarily correct \cite{brown2020language, radford2019language}. People often misinterpret fluency as a signal of truth \cite{ghafouri2024epistemic}, which leads to premature acceptance of initial outputs \cite{bhole2021dangers}. Over time, frequent exposure to confident and consistent responses generates familiarity, which users mistake for accuracy. The model’s stylistic consistency becomes a heuristic for epistemic reliability.

With repetition, AI systems are reclassified from instruments to epistemic partners. Automation bias and the cultural association of computation with objectivity amplify this perception \cite{crockett2025modern, hutson2025godprompt, Craik1952, Smyth1994}. As models improve, their language acquires an aura of omniscience reinforced by cultural metaphors that equate knowledge with divinity \cite{Liang2022, Tejeda2022, Kelly2023, Karatas2023}. Once established, this trust anchors subsequent reasoning: the first output provided by the model shapes users’ expectations and constrains their search for alternatives \cite{Jansson1991, Purcell1996, Crilly2019, Gaver2003, Sundar2019}. Within collective environments, these anchors are magnified by conformity pressures, as users converge on the statistical mean of machine-generated consensus \cite{cialdini2009influence, weidinger2021ethical}.

AI-derivative epistemology therefore connects microcognitive tendencies to macrolevel outcomes. Individual acts of deference accumulate into collective homogenization. What begins as a technical optimization for predictive accuracy becomes, through human trust and repetition, a social mechanism that narrows epistemic diversity. This process completes the circuit of the AI Prism: the model compresses informational variance, and human cognition scales that compression into the structure of collective knowledge.

\begin{figure}[t]
\centering
\resizebox{\linewidth}{!}{%
\begin{tikzpicture}[node distance=2.4cm, every node/.style={font=\small}, >=Latex]
\tikzstyle{box}=[draw, rounded corners, align=center, inner sep=4pt, text width=3.8cm]

\node[box] (drivers) {Technical and human\\ drivers of convergence};
\node[box, right=2.6cm of drivers] (similar) {Outputs become\\ stylistically and conceptually similar};
\node[box, right=2.6cm of similar] (feedback) {Feedback and reuse\\ amplify convergence};

\draw[->] (drivers) -- (similar);
\draw[->] (similar) -- (feedback);

\node[box, below=2.8cm of drivers] (translate) {Translation and\\ simplification of ideas};
\node[box, right=2.6cm of translate] (recombine) {Human recombination\\ across domains};
\node[box, right=2.6cm of recombine] (novelty) {New combinations\\ and innovations};

\draw[->] (translate) -- (recombine);
\draw[->] (recombine) -- (novelty);

\draw[dashed,->] (similar.south) .. controls +(0,-1.4) and +(0,1.4) .. node[midway,right]{simplified ideas as building blocks} (recombine.north);

\node[draw, rounded corners, fit=(drivers)(feedback)(translate)(novelty), inner sep=10pt] {};
\end{tikzpicture}
}
\caption{\small The Prism–Bridge framework. The upper pathway shows how technical optimization and human deference narrow informational diversity. The lower pathway shows how these standardized forms become portable and enable new combinations across domains. The dashed link represents how homogenization provides the raw material for recombination.}
\label{fig:prism-bridge}
\end{figure}

\subsection*{System-level consequences}

The broader impact of the AI Prism lies not in model architecture alone but in its diffusion through society. Informational variance is the foundation of adaptation and innovation \cite{Henrich2004, Hong2004, Muthukrishna2016}. When the compression of language at the model level is combined with widespread epistemic deference at the user level, local reductions in diversity scale into systemic homogenization. These two forces—technical optimization and social uptake—jointly reorganize how knowledge is produced and circulated.

As generative systems become embedded across institutions, a growing share of communication is mediated by the same few foundation models \cite{Messeri2024}. This shared infrastructure promotes stylistic and conceptual convergence. Studies already document measurable declines in lexical and semantic diversity in AI-assisted writing \cite{Wu2024, Leppanen2025, sourati2025shrinking, Angelidis2021}. The result is a narrowing of expressive repertoires that constrains collective search and limits the emergence of new interpretive frames \cite{Foucault1969, gadamer1975hermeneutics, ricoeur1981hermeneutics, Fazelpour2025}.

Homogenization is reinforced by changes in human practice. Reliance on generative tools replaces forms of reasoning that once depended on exploratory effort. Historical evidence shows that dependence on external cognitive aids can erode skills in the absence of compensatory training \cite{carr2020shallows, heersmink2016internet, sparrow2011google}. Generative AI extends this tendency from mechanical to conceptual domains. When systems supply complete drafts or structured arguments, users may lose not only procedural fluency but also the heuristics required for critical reflection and dissent \cite{Doshi2024, castro2023human}. As human cognition adapts to these tools, the intellectual environment becomes more efficient yet less generative.

The cumulative effect is a recursive narrowing of the informational ecosystem. As model outputs are integrated back into training data, distributions begin to converge around their own synthetic averages. Empirical studies demonstrate that such feedback accelerates a degenerative process known as model collapse, where variance diminishes rapidly as synthetic data dominates the training corpus \cite{razeghi2022impact, Thiele2025b, shumailov2024ai, Qiu2025b}. This recursive dynamic transforms compression from a technical property into a structural condition of communication.

Taken together, these developments define the system-level manifestation of the AI Prism. The convergence of technical optimization, epistemic deference, and recursive feedback produces informational environments that are increasingly coherent yet progressively less diverse. Efficiency rises at the cost of adaptive flexibility. The same forces that make communication seamless also constrain the irregularities that sustain discovery. Yet this is only half of the paradox. The reduction of local variance also establishes the preconditions for recombination and the emergence of global novelty.

\begin{tcolorbox}[floatplacement=!ht, float,
    title=\textbf{Box 2 | Conceptual Glossary},
    colback=white,
    colframe=black,
    fonttitle=\bfseries,
    sharp corners,
]
\begin{tabularx}{\linewidth}{@{} p{3.5cm} X @{}}
\textbf{Term} & \textbf{Definition} \\
\midrule
\addlinespace 

Generative AI & A class of artificial intelligence systems that produce new outputs (e.g., text, images, code) by learning statistical patterns from large datasets. Large Language Models (LLMs) are the most prominent example. \\
\addlinespace

Epistemic & Pertaining to knowledge: its production, justification, and circulation. “Epistemic” refers to both individual beliefs and the collective conditions under which knowledge is generated, validated, and diffused. \\
\addlinespace

Homogenization \& Variance Compression & The reduction of informational variance across AI outputs, understood here as stylistic, semantic, and conceptual convergence in generated content. Statistically, this corresponds to a narrowing of the distribution of possible outputs. \\
\addlinespace

AI Prism & A socio-technical process through which generative AI reduces informational variance by compressing distributions, aligning outputs with user preferences, and diffusing them through centralized infrastructures. \\
\addlinespace

AI-Derivative Epistemology & A usage pattern in which users delegate synthesis/evaluation to an AI and treat its output as primary evidence. This deference enables variance compression via the AI Prism. \\
\addlinespace

Recombinant Innovation & The creation of novelty through the recombination of existing elements. Emphasized in economics, psychology, and cultural evolution as the dominant mechanism of innovation. \\
\addlinespace

Paradoxical Bridge & The mechanism by which homogenization enables novelty. Standardized outputs from AI travel across domains, lowering translation costs and increasing the probability of cross-domain recombination. \\
\addlinespace

Liquefaction & The semantic process by which AI systems translate domain-specific representations into fluid and portable forms, increasing translation between knowledge domains. \\
\addlinespace
\end{tabularx}
\end{tcolorbox}

\section*{Paradoxical Bridge}
Paradoxically, generative artificial intelligence compresses human expression while expanding the space of creative possibility. Its predictive optimization narrows linguistic and conceptual variance \cite{Anderson2024, Prakash2025, Leppanen2025, sourati2025shrinking, Boden2004, Zhou2024, Vaithilingam2022, sarkar2023exploring, Homogenizing2025, Bender2021}, yet under specific conditions this same compression produces the reverse effect \cite{gao2023quantifying, LeeChung2024, Shiiku2025, Krakowski2025, Johansen2025, zhou2025expands}. We call this inversion the Paradoxical Bridge, the process through which informational convergence becomes the foundation for large-scale recombination.

As generative models compress variation, they make knowledge modular and transferable across domains. Compression liquefies meaning \cite{galison1997image, luhmann1995social}. Specialized vocabularies are translated into shared representational forms that move easily between contexts. When disciplinary languages converge, translation costs fall and concepts circulate across previously separated fields. Ideas that once required specialized expertise become accessible to broader communities of practice.

The Paradoxical Bridge explains how standardization can enable novelty. The effect depends on human agency. When users treat standardized outputs as raw material rather than final products, the same forces that homogenize expression within domains generate diversity across them. In this view, generative systems create the structural conditions for recombinant innovation, a pattern documented in computational studies of cross-domain synthesis and creative transfer \cite{guzdial2018combinets, gu2024llms, choi2024creativeconnect, radensky2024scideator, guo2025streamlining, tomita2025extracting, sternlicht2025chimera, zhao2025ramon}.

\subsection*{Innovation and recombination}
The Paradoxical Bridge rests on a particular conception of innovation. Rather than treating novelty as incremental improvement \cite{price1963little} or the product of individual inspiration \cite{galton1891hereditary, barron1975solitariness}, we follow a long tradition in economics, psychology, and cultural evolution that defines innovation as recombination. Creativity arises from rearranging existing elements into new configurations \cite{Campbell1960, sneed1987scientific, boden2009computer, Simonton2010, gluaveanu2014distributed, Muthukrishna2016}. Innovation does not emerge from the void but from the reorganization of what already exists. This definition links diverse accounts of change in science, technology, and culture under a shared conceptual logic.

A large body of evidence supports this perspective. Einstein described his insights as “combinatorial play” \cite{hadamard1945psychology}. Schumpeter argued that economic progress depends on “new combinations” that disrupt equilibrium \cite{Schumpeter1942}. Contemporary studies point in the same direction. High-impact research tends to integrate previously unconnected domains \cite{Uzzi2013, Kim2016, thagard2012creative}. Breakthrough patents often emerge from the fusion of distant technologies. Institutions that encourage cross-domain collaboration experience faster cycles of discovery and innovation \cite{mokyr2011gifts, furman2011climbing, goodman2025institutional}. Recent computational work extends this principle to artificial intelligence itself, showing that both human and machine creativity increase when source materials are diverse and domain boundaries are porous \cite{gao2024cross, muller2025digital}.

Formal models of recombinant growth capture the same idea \cite{Weitzman1998, Arthur2009, Padgett2012, agrawal2018finding, shin2023superhuman, brinkmann2023machine}. They describe innovation as a search through combinatorial space where the likelihood of discovery depends on the connections among existing components. Generative AI accelerates this search by liquefying and translating knowledge across fields. This process enlarges the space of possible combinations and raises the probability of cross-domain synthesis. In this sense, AI functions as a catalyst for recombination.

\subsection*{AI Innovation and Recombination}
Generative AI turns local homogenization into a foundation for global novelty \cite{boden1998creativity}. The process begins with meaning liquefaction \cite{sarkar2022like}. By standardizing outputs, generative models act as universal translators that render specialized concepts into accessible and interoperable forms \cite{dunnell2023latent, wong2025llm2tea}. As disciplinary boundaries dissolve, expertise that was once confined to specific domains can circulate more freely. This circulation broadens the range of actors who can engage with technical material and reconfigure it in new contexts, a pattern long recognized in studies of discovery and knowledge transfer \cite{bao2025words}. In mathematics, for instance, systems trained jointly on proofs, code, and natural language have merged symbolic reasoning into shared representational spaces, allowing conjectures that connect distant subfields such as topology and algebra.

Liquefaction expands access but also produces cognitive strain. A flood of cross-domain material overwhelms human attention, creating a bottleneck where ideas compete for limited cognitive resources \cite{sarica2024innovation}. The next stage, cognitive offloading, addresses this constraint. By delegating filtering, summarization, and compression to AI, users relieve themselves of lower-level assimilation tasks \cite{lubart2005can, holgersson2024open}. The reduction in cognitive load enables focus on higher-order reasoning such as comparison, abstraction, and the detection of latent relationships between domains.

Novelty returns only when human curation is introduced into the loop. Here, the user actively directs the generative process rather than accepting it passively. Curation involves setting goals, applying evaluation criteria, and integrating insights across multiple outputs. It draws on a combination of domain expertise, cognitive flexibility, and epistemic humility, which together allow users to identify promising connections that the model itself cannot infer \cite{mollick2024co, peschl2024human, grilli2024creativity, he2024creative, chamakiotis2024unpacking, zhu2025ai, rozental2025artists, pedota2025human}. The Bridge depends on this final human intervention. Without it, compression remains an end state rather than a platform for novelty \cite{hossbach2025ai, bordas2025switching, shen2025understanding, chiriatti2025system}.

The dynamics of liquefaction, cognitive offloading and human curation scale beyond individuals to important systemic changes. As meaning becomes liquefied and cognitive work redistributed, the production of knowledge reorganizes itself around two linked transformations. The first is a broad democratization of innovation. Lower translation costs open the field of creative recombination to a wider range of actors \cite{Garcia2024}. Ideas that were once restricted to specialists become accessible to generalists and outsiders who can reinterpret them in unfamiliar contexts. Historical analyses of discovery show that breakthroughs often emerge from these margins, where unconventional perspectives expose relationships that experts overlook \cite{merton1973sociology, fleming2007collaborative, burt2004structural}. While a more inclusive recombination process may produce uneven quality in the short term, over time it enlarges the combinatorial search space and supplies the system with a richer set of conceptual building blocks.

As standardized forms circulate, the geography of creativity and variance shifts. Homogenization first consolidates local knowledge by creating standardized and interoperable forms within disciplines. Once these forms circulate, they acquire new mobility. Concepts stripped of disciplinary jargon can cross boundaries, linking previously isolated communities. What begins as a decline in local diversity becomes a precondition for global diversity. As standardized elements recombine across domains, informational variance increases at the system level. This inversion follows a long-observed pattern in cultural and scientific evolution, where phases of consolidation precede bursts of innovation \cite{Schumpeter1942, Kuhn1962, Boyd1985, Mesoudi2011, Henrich2016}. The terrain of creativity shifts from within established fields to the networked space between them.

Together, these transformations describe how the Bridge couples order with renewal. The system trades local specialization for global connectivity. The relationship between the Prism and the Bridge follows a U-shaped temporal dynamic. Early in the diffusion of generative AI, convergence dominates and informational variance declines. As standardized representations proliferate and circulate, a second phase of recombinant novelty emerges. The health of the knowledge ecosystem depends on sustaining this equilibrium—preserving enough uniformity for translation while maintaining sufficient heterogeneity for discovery across scientific, technological, and cultural domains.

\section*{Empirical and Policy Implications}
To make the Prism–Bridge framework empirically tractable, its mechanisms must link measurable changes in variance compression to subsequent creative recombination. This presents two primary empirical challenges. The first is causal clarity. Since novelty and variance are recursively connected, initial tests must isolate first-order effects (changes within a fixed representational space) from second-order effects (where novelty reshapes the space itself) to avoid endogeneity.

The second challenge is the level of analysis. While our framework concerns system-level dynamics, its mechanisms must be observed at the micro-level. Following established practice in cultural evolution, empirical analysis should begin with individuals and teams as observable units \cite{Henrich2004, Mesoudi2011, Muthukrishna2016}. Aggregating the behavior of these units is what produces the macro-level U-shaped dynamic our framework predicts, allowing for a measurable bridge between micro-level observation and macro-level theory.

\begin{figure}[h!]
\centering
\begin{tcolorbox}[
    boxrule=0.5pt,
    arc=2pt,
    sharp corners,
    boxsep=2mm, 
    colback=white, colframe=black,
    width=\linewidth 
]
\centering 

\begin{tikzpicture}[
    >=Latex,
    node distance=1.0cm and 4.5cm, 
    scale=0.75,                  
    transform shape
]

\tikzstyle{box}=[draw, rounded corners, align=center, text width=3.3cm, minimum height=1.5cm] 
\tikzstyle{modbox}=[draw, rounded corners=3pt, align=center, text width=3.3cm, minimum height=0.8cm]

\node[box] (T) 
{\textbf{Model Design (1) }\\[2pt]\footnotesize(e.g., training data, interface)};

\node[box, below=of T] (ADE) 
{\textbf{AI-Derivative Epistemology (2)}\\[2pt]\footnotesize(User's psychological stance)};

\coordinate (helper) at ($(T.east)!0.5!(ADE.east)$);

\node[box, right=of helper] (M) 
{\textbf{Prism: Variance Compression (3)}\\[2pt]\footnotesize(Informational diversity)};

\node[box, right=of M] (Y) 
{\textbf{Bridge: Recombination (4)}\\[2pt]\footnotesize(Novelty outcomes)};

\node[modbox, below=of M, yshift=0.5cm] (C) 
{\textbf{Contextual Moderators}\\[2pt]\footnotesize(e.g., domain, expertise, time)};

\draw[->, thick] (T) -- (ADE);
\draw[->, thick] (helper) -- (M); 

\draw[->, thick] (C) to [bend right=15] (ADE);
\draw[->, thick] (C) -- (M);

\path (Y.west) coordinate (Yleft);

\draw[->, thick] (M) -- ($(Yleft)+(-0.2,0)$);
\draw[->, thick] (C) to[bend left=15] ($(Yleft)+(-0.2,0)$);

\end{tikzpicture}
\end{tcolorbox}
\caption{
    \small The Causal Architecture of the Prism–Bridge Framework. Model design influences the adoption of a user's AI-derivative epistemology. The interaction of these two factors jointly causes the Prism (variance compression), which in turn determines the potential for novelty recombination. This entire process is moderated by contextual factors.
}
\label{fig:dag}
\end{figure}

\subsubsection*{Mapping the Causal Pathway}
Figure~\ref{fig:dag} summarizes the core causal pathway of the framework. This is a four-step process. Variation in (1) Generative Model Design (e.g., training data, interface) influences the adoption of (2) AI-Derivative Epistemology, the psychological stance of user deference. The interaction between (1) and (2) is the driver of (3) Variance Compression (the Prism), where informational diversity is reduced. Finally, the level of variance in the information environment mediates downstream effects on (4) Creative Recombination (the Bridge). Contextual factors such as domain conventions and time moderate these relationships at multiple stages. Table~\ref{tab:ops} lists representative measures for each stage of the pathway.

\begin{table}[h!]
\centering
\caption{Operationalizing the Prism–Bridge Causal Pathway}
\label{tab:ops}
\small
\begin{tabularx}{\linewidth}{@{} X X X X @{}} 
\toprule
\textbf{Model Design} & \textbf{AI-Derivative Epistemology} & \textbf{Homogenization} & \textbf{Recombination} \\
\midrule
Diversity of training data & Low modification rate of outputs & Increased textual similarity across users & Cross-domain citation or concept borrowing \\

Alignment protocol & High acceptance rate of AI suggestions & Decline in lexical and syntactic diversity & Emergence of hybrid genres or styles \\

Interface design and affordances & Reduced self-reported cognitive effort & Standardization of formatting and structure & Integration of components from distinct modules \\

Decoding parameters (e.g., temperature) & High reliance on AI for evaluation & Concentration of outputs in narrow conceptual space & Increased co-occurrence of previously distant topics \\
\bottomrule
\end{tabularx}
\end{table}

\subsubsection*{Empirical Designs and Identification Strategies}

Testing the Prism–Bridge framework requires tracking how changes in system design affect informational diversity and, subsequently, creative recombination. Table~\ref{tab:designs} summarizes empirical strategies that can identify this causal sequence. Since diversity and novelty co-evolve, identification must rely on exogenous or time-lagged interventions that modify dispersion without directly determining recombination.

\vspace{0.5cm}
\begin{table*}[t]
\centering
\caption{Empirical strategies for testing the Prism–Bridge framework}
\label{tab:designs}
\small
\begin{tabular}{p{0.18\linewidth} p{0.32\linewidth} p{0.45\linewidth}}
\toprule
\textbf{Design} & \textbf{Setting} & \textbf{Purpose} \\
\midrule
Controlled experiment & Manipulate generative system design parameters such as prompt structure, training context, temperature, or interface cues & Identifies the causal link between variance compression and recombination efficiency under controlled conditions \\[4pt]

Field quasi-experiment & Natural or policy-induced changes in AI tools used by firms, schools, or media platforms & Observes whether shifts toward standardization affect downstream novelty, collaboration, or diffusion patterns \\[4pt]

Platform A/B intervention & Deploy diversity-promoting or contrastive AI assistants in online collaboration environments & Tests whether interventions that reintroduce variance increase cross-domain linkages or hybrid outputs \\[4pt]

Archival longitudinal study & Track variance and novelty metrics before and after model diffusion  & Estimates the temporal lag between homogenization and innovation at the system level \\[2pt]
\bottomrule
\end{tabular}
\end{table*}
\vspace{0.5cm}

\subsubsection*{Predictions and Future Directions}

The Prism–Bridge framework's U-shaped temporal dynamic yields several falsifiable claims, summarized in Table~\ref{tab:predictions}. The central prediction is a two-phase temporal sequence. Phase one, dominated by the Prism, should show a negative correlation between AI adoption and informational variance. Phase two, where the Bridge becomes functional, should show a positive correlation between the newly established "standardized forms" and the rate of recombinant novelty.

This dynamic has an identifiable temporal shape. The framework predicts that an initial decline in novelty will be followed by a measurable delay, after which recombinant novelty should begin to increase \cite{solow1957technical, brynjolfsson2014second, brynjolfsson2021productivity, tutuncuoglu2024beyond}. The magnitude of this delay will depend on domain-specific factors such as model diffusion rates and adoption cycles, but its existence is a core test of our proposed causal order. Future research should focus on refining diversity and novelty metrics for longitudinal analysis, replicating these dynamics across domains, and evaluating interventions designed to shorten the initial dip and accelerate subsequent recovery.

\vspace{0.5cm}
\begin{table}[h!]
\centering
\caption{Testable Predictions of the U-Shaped Dynamic}
\label{tab:predictions}
\small
\begin{tabular}{p{0.3\linewidth} p{0.65\linewidth}}
\toprule
\textbf{Prediction} & \textbf{Empirical Implication} \\
\midrule
\addlinespace
\textbf{Mediation Effect} & Design and user stance jointly determine informational variance, which in turn governs the potential for recombinant novelty. Statistical tests should reveal variance as the intermediate variable linking design to creative output. \\
\addlinespace
\textbf{U-Shape} & Longitudinal analysis of a domain after widespread AI adoption should reveal an initial \textit{decrease} in novelty/variance, followed by a subsequent \textit{increase} in recombinant novelty. \\
\addlinespace
\textbf{Temporal Lag} & The rise in recombinant novelty will not be immediate. It will appear only after a measurable delay following the initial period of convergence and standardization. \\
\addlinespace
\textbf{Human Curation is Key} & The U-shaped recovery (the upward swing) is conditional on active human curation. In contexts of purely automated or uncritical use, novelty will decline and fail to recover. \\
\addlinespace
\textbf{Falsifiability Condition} & The framework is falsified if, after widespread AI adoption, novelty either (a) declines monotonically with no subsequent recovery, or (b) increases immediately with no initial dip. \\
\bottomrule
\end{tabular}
\end{table}
\vspace{0.5cm}

\subsection*{Governance Implications}
The Prism–Bridge framework reframes the purpose of AI governance. The task is not to maximize diversity at all times, as some proposals suggest \cite{sourati2025shrinking}, but to manage the U-shaped dynamic of informational variance, softening the initial compression while fostering the conditions for recombinant recovery.

Governance should shorten the homogenizing phase by increasing the recombinability of standardized outputs. This can be achieved through incentives for architectural diversity, investment in interoperable infrastructures, and support for open evaluation systems that sustain multiple model lineages.

The key governance challenge is to ensure the upward swing of the U-curve occurs. Because recovery depends on human agency, policy must cultivate curatorial capacity rather than content production. Education must shift from teaching content generation to teaching the critical capacities of an expert curator: how to iteratively prompt, critically evaluate, and imaginatively reconfigure AI outputs \cite{distefano2025evaluating, luchini2025roles, zhang2025can}. In parallel, technical design should create tools that encourage comparison and synthesis rather than passive deference to AI-derivative epistemology \cite{yatani2024ai, schulzeyegpt2025, kim2025steering}. The goal is to build the human capital required for the recombinant phase to ignite.

A temporary dip in variance can be productive, but a full collapse would be catastrophic. Governance must preserve baseline heterogeneity in data, models, and linguistic representation to prevent irreversible lock-in. Measures such as heritage data vaults and pluralistic model ecosystems ensure enough raw material persists for future recombination \cite{jenkins2024principle, frosio2025should, lucchi2025generative}.

\begin{tcolorbox}[
    title=\textbf{Box 3 | Scope, Limitations, and Critiques},
    colback=white,
    colframe=black,
    fonttitle=\bfseries,
    sharp corners,
    before skip=\baselineskip,
    after skip=\baselineskip,
    fontupper=\footnotesize
]

\begin{itemize}\itemsep0.75em
    \item \textbf{Technological Context:} The \textit{Prism}'s homogenizing force is strongest for today’s large, centralized models that favor stability over variance \cite{he2025human}. A more decentralized or pluralistic model ecosystem would alter these dynamics.

    \item \textbf{User Behavior:} The framework assumes a general tendency toward deference, especially under pressure. If AI increasingly displaces rather than complements human judgment, the bridge could erode, and homogenization would fail to be recombinable.

    \item \textbf{Non-Linearity:} The relationship is not linear. Homogenization both enables and constrains. Excessive convergence destroys the very fragments needed for recombination, meaning moderate standardization fosters novelty while extremes collapse it.
    
    \item \textbf{Countervailing Forces:} Real-world systems contain damping forces that limit total convergence, such as market pressures for differentiation, cultural drift, and institutional diversity. These mechanisms may ensure homogenization remains partial and reversible.

    \item \textbf{Domain sensitivity}: The framework predicts a general U-shaped dynamic, but its specific shape and timing will likely vary by domain. We expect a slower recovery in fields with rigid evaluative norms and high costs for error, such as medicine or legal practice. In contrast, fields that are more tolerant of hybridization and place a higher value on novelty, such as the visual arts, academic work, and entrepreneurship, may experience a more pronounced upward swing. Future empirical work must therefore account for the specific ecology of each knowledge domain.
    
    \item \textbf{Shallow Synthesis vs. Deep Innovation:} Standardized forms may yield only superficial "shallow" novelty. While history shows even shallow transfers can catalyze deep innovation \cite{Taalbi2018, Wang2023, Jain2003, Brooks2021, Bernhofen2016}, productive recombination must ultimately balance novelty with usefulness and coherence \cite{Simonton1999, Boden2004}.

    \item \textbf{Lack of ideas to recombine}: Compression reduces the visible diversity of ideas available for recombination. Yet the global reservoir of knowledge remains vastly larger than any individual or institution can navigate. Even standardized forms preserve immense latent variety. The real constraint is not a shortage of variation but the difficulty of navigating abundance—a condition under which structured compression can enhance, rather than hinder, creative synthesis.

    \item \textbf{Mechanism Clarification (vs. Retrieval):} The framework's core mechanism is distinct from efficient information retrieval. Retrieval filters and ranks existing information. Homogenization, by contrast, alters the architecture of knowledge itself by standardizing it for portability. This enables deep integration, whereas retrieval alone may only yield juxtaposition. The \textit{Prism} and \textit{Bridge} are therefore best understood as dominant tendencies within the current generative technological context.
\end{itemize}
\end{tcolorbox}
\vspace{1cm}

\section*{Final Remarks}
This article has developed a framework for the dialectic between homogenization and recombination in the age of generative AI. We have argued that the AI Prism drives homogenization by compressing variance through machine optimization and human deference, while the Paradoxical Bridge can enable novelty, as those standardized forms circulate across domains and become inputs for recombination.

The implications of this dialectic extend across domains. In science, the Prism may standardize outputs while the Bridge lowers barriers to interdisciplinary collaboration. In culture, homogenization can compress stylistic diversity even as recombination generates new hybrids. In political economy, these same forces can either entrench incumbents or empower boundary-crossers, depending on institutional design. Across each of these domains, long-term resilience depends not on maximizing one force, but on intelligently balancing them.

The contribution of this framework extends beyond artificial intelligence. It offers a general account of how knowledge evolves under the dual pressures of consolidation and renewal. Parallel dynamics of stabilization followed by disruptive recombination have long been observed across the social and historical sciences. By formalizing these tendencies as an interdependent socio-technical process, our framework reframes current debates. The central question for the future is not whether AI will increase productivity or erode diversity, but how the balance between the Prism and the Bridge will be understood, managed, and governed.

\newpage
\nolinenumbers
\printbibliography

\immediate\write18{texcount -inc -sum -nobib -v0 \
  -cs=cite,cite,ignore \
  -cs=textcite,textcite,ignore \
  -cs=cite,cite,ignore \
  main.tex > count.txt}

\end{document}